\documentclass[twocolumn,prl,aps,amsmath,amssymb,epsfig]{revtex4}

\usepackage{graphicx}% Include figure files
\usepackage{dcolumn}% Align table columns on decimal point
\usepackage{bm}% bold math

\begin{document}
\title{Confinement versus Chiral Symmetry}
 \author{\'{A}gnes {\sc M\'{o}csy}}
 \email{agnes.mocsy@nbi.dk}
 \author{Francesco {\sc Sannino}}
 \email{francesco.sannino@nbi.dk}
 \author{Kimmo {\sc Tuominen}}\email{tuominen@nordita.dk}
 \affiliation{The Niels Bohr Institute \& {\rm NORDITA}, Blegdamsvej
  17, DK-2100 Copenhagen \O, Denmark }
\date{August 2003}

\begin{abstract}
We construct an effective Lagrangian which illustrates why color
deconfines when chiral symmetry is restored in hot gauge theories
with quarks in the fundamental representation. For quarks in the
adjoint representation we show that while deconfinement and the
chiral transition do not need to coincide, entanglement between
them is still present. Extension to the chemical potential driven
transition is discussed.
\end{abstract}

\maketitle

In the absence of quarks the $SU(N)$ Yang-Mills theory has a
global $Z_N$ symmetry \cite{Svetitsky:1982gs}. There exists a
gauge invariant operator charged under $Z_N$, the Polyakov loop,
which can be identified as the order parameter of the theory, and
thus be used to characterize the deconfinement phase transition
\cite{Polyakov:vu}. One can directly study this phase transition
via numerical lattice simulations. Such studies have revealed that
the deconfinement phase transition is second order when the number
of colors is $N_{\rm c}=2$ \cite{Damgaard:1987wh}, weakly though
\cite{Kaczmarek:1999mm}, but first order for $N_{\rm c}=3$
\cite{Bacilieri:1988yq}, and presumably first order for $N_{\rm
c}\geq 4$ \cite{Batrouni:1984vd}.

The picture changes considerably when quarks are added to the
theory. If fermions are in the fundamental and pseudoreal
representations for $N_{\rm c}=3$ and $N_{\rm c}=2$, respectively,
the corresponding $Z_3$ or $Z_2$ center of the group is never a
good symmetry. The order parameter is the chiral condensate which
characterizes the chiral phase transition. For $N_c$=3 and two
massless quark flavors at finite temperature and zero baryon
density, the chiral phase transition is in the same universality
class as the three dimensional $O(4)$ spin model
\cite{Wilczek:1992sf}, becoming a smooth crossover as small quark
masses are accounted for \cite{Scavenius:2000qd}. For $N_c=2$ the
relevant universality class is that of $O(6)$ both for the
fundamental and adjoint representations \cite{Holtmann:2003he}.
Even if the discrete symmetry is broken, one can still construct
the Polyakov loop and study the temperature dependence of its
properties on the lattice. One still observes a rise of the
Polyakov loop from low to high temperatures and naturally,
although improperly, one speaks of deconfining phase transition
\cite{Karsch:1998qj}. For fermions in the adjoint representation
the center of the group remains a symmetry of the theory, and thus
besides the chiral condensate, also the Polyakov loop is an order
parameter.

Interestingly, lattice results \cite{Karsch:1998qj} indicate that
for ordinary QCD with quarks in the fundamental representation,
chiral symmetry breaking and confinement (i.e. a decrease of the
Polyakov loop) occur at the same critical temperature. Lattice
simulations also indicate that these two transitions do not happen
simultaneously when the quarks are in the adjoint representation.
Despite the attempts to explain these behaviors \cite{Brown:dm},
the underlying reasons are still unknown.

In this Letter we propose a solution to this puzzle based on the
approach presented in \cite{Mocsy:2003tr,{Mocsy:2003un}},
envisioned first in \cite{Sannino:2002wb}, concerning the transfer
of critical properties from true order parameters to non-critical
fields. The order parameter field is a field whose expectation
value is a true order parameter, i.e. is zero in the symmetric
phase and non-zero in the spontaneously broken one. The non-order
parameter (or non-critical) fields are the ones whose expectation
values do not have such a behavior.

Two general features introduced in
\cite{Mocsy:2003tr,{Mocsy:2003un}} are essential: There exists a
relevant trilinear interaction between the light order parameter
and the heavy non-order parameter field, singlet under the
symmetries of the order parameter field. This allows for an
efficient transfer of information from the order parameter to the
fields that are singlets with respect to the symmetry of the
theory. As a result, the non-critical fields have infrared
dominated spatial correlators. The second feature, also due to the
existence of such an interaction, is that the finite expectation
value of the order parameter field in the symmetry broken phase
induces a variation in the expectation value for the singlet
field, whose value generally is non-vanishing in the unbroken
phase.

%%%%%%%%%%%%%%%%%%%%%%%%%%%%%%%%%%%%
\section{Fundamental Representation}
\label{fundamental}

Here we study the behavior of the Polyakov loop by treating it as
a heavy field that is a singlet under chiral symmetry
transformations. We take the underlying theory to be two colors
and two flavors in the fundamental representation. The degrees of
freedom in the chiral sector of the effective theory are
$2N_f^2-N_f-1$ Goldstone fields $\pi^a$ and a scalar field
$\sigma$. For $N_f=2$ the potential is
\cite{Appelquist:1999dq,Lenaghan:2001sd}:
\begin{eqnarray}
V_{\rm ch}[\sigma,\pi^a]&=&\frac{m^2}{2}{\rm Tr
}\left[M^{\dagger}M\right]+ {\lambda_1}{\rm Tr
}\left[M^{\dagger}M\right]^2\nonumber \\&+&
\frac{\lambda_2}{4}{\rm Tr }\left[M^{\dagger}MM^{\dagger}M\right]
\label{chiralpot}
\end{eqnarray}
with $2\,M=\sigma + i\,2\sqrt{2}\pi^a\,X^a$, $a=1,\dots,5$ and
$X^a\in {\cal A}(SU(4))-{\cal A}(Sp(4))$. $X^a$ are the generators
provided explicitly in equation (A.5) and (A.6) of
\cite{Appelquist:1999dq}. The Polyakov loop potential in the
absence of the $Z_2$ symmetry is
\begin{eqnarray}
V_\chi[\chi]=g_0\chi+\frac{m_\chi^2}{2}\chi^2+\frac{g_3}{3}\chi^3
+\frac{g_4}{4}\chi^4 \, . \label{chipot}
\end{eqnarray}
The field $\chi$ represents the Polyakov loop itself, while
$m_{\chi}$ is the mass above the chiral phase transition. To
complete the effective theory we introduce interaction terms
allowed by the chiral symmetry
\begin{eqnarray}
V_{\rm{int}}[\chi,\sigma,\pi^a]&=& \left(g_1\chi +g_2\chi^2\right){\rm Tr }
\left[M^{\dagger}M\right] \nonumber \\ &=&
\left(g_1\chi +g_2\chi^2\right)(\sigma^2+\pi^a\pi^a) \, .
\end{eqnarray}
In the phase with $T<T_{c\sigma}$, where chiral symmetry is
spontaneously broken, $\sigma$ acquires a nonzero expectation
value, which in turn induces a modification also for
$\langle\chi\rangle$. The usual choice for vacuum alignement is in
the $\sigma$ direction, i.e. $\langle\pi\rangle=0$. %Assuming small
%fluctuations of the heavy $\chi$ field,
The extremum of the linearized potential is at
\begin{eqnarray}
\langle\sigma\rangle^2 &\simeq& -\frac{m^2_{\sigma}}{\lambda}\, ,
\qquad \quad\,\,\, \quad m^2_{\sigma} \simeq m^2 + 2g_1\langle
\chi\rangle \label{vevsigma} , \\ \langle \chi \rangle\,\,
&\simeq& \chi_0 -\frac{g_1}{m_\chi^2}\langle\sigma\rangle^2\, ,
\quad ~ \chi_0 \simeq -\frac{g_0}{m^2_{\chi}} \ ,
 \label{vevchi}
\end{eqnarray}
where $\lambda=\lambda_1 + \lambda_2$. Here $m^2_{\sigma}$ is the
full coefficient of the $\sigma^2$ term in the tree-level
Lagrangian which, due to the coupling between $\chi$ and $\sigma$,
also depends on $\langle\chi\rangle$. Spontaneous chiral symmetry
breaking appears for $m^2_{\sigma} <0$. In this regime the
positive mass squared of the $\sigma$ is $M^2_{\sigma} = 2
\lambda\langle \sigma ^2 \rangle$. The formulae (\ref{vevsigma})
and (\ref{vevchi}) hold near the phase transition where $\langle
\sigma \rangle$ is small. We have ordered the couplings such that
$g_0/m_{\chi}^3$ and $g_1/m_{\chi}$ are both much greater than $
g_2$ and $g_3/m_{\chi}$. This previous ordering does not affect
our general conclusions. {}No such ordering will be considered for
quarks in the adjoint representation of the gauge group. When
computing the expectation values for the relevant fields we will
keep the full potential.

Near the critical temperature the mass of the order parameter
field is assumed to posses
%, for continuous phase transitions,
the generic behavior $m_\sigma^2\sim (T-T_{\rm{c}})^\nu$. Equation
(\ref{vevchi}) shows that for $g_1>0$ and $g_0<0$ the expectation
value of $\chi$ behaves oppositely to that of $\sigma~$: As the
chiral condensate starts to decrease towards chiral symmetry
restoration, the expectation value of the Polyakov loop starts to
increase, signaling the onset of deconfinement. This is
illustrated in the left panel of figure \ref{Figura1}. Positivity
of the expectation values implies $2g_1^2-\lambda m_\chi^2<0$,
which also makes the extremum a minimum. At the one-loop level one
can show \cite{Mocsy:2003un} that also $\chi_0$ acquires a
temperature dependence.

When applying the analysis presented in
\cite{Mocsy:2003tr,{Mocsy:2003un}}, the general behavior of the
spatial two-point correlator of the Polyakov loop can be obtained.
Near the transition point, in the broken phase, the $\chi$
two-point function is dominated by the infrared divergent
$\sigma$-loop. This is so, because the $\pi^a$ Goldstone fields
couple only derivatively to $\chi$, and thus decouple. We find a
drop in the screening mass of the Polyakov loop at the phase
transition. When approaching the transition from the unbroken
phase the Goldstone fields do not decouple, but follow the
$\sigma$, resulting again in the drop of the screening mass of the
Polyakov loop close to the phase transition. We consider the
variation $\Delta m_\chi^2(T)=m_\chi^2(T)-m_\chi^2$ of the $\chi$
mass near the phase transition with respect to the tree level mass
$m_{\chi}$. The one loop analysis predicts:
\begin{eqnarray}
\Delta m^2_\chi(T)&\sim& - \frac{g_1^2}{|m_{\sigma}|}\sim t^{-\frac{\nu}{2}} ,
\end{eqnarray}
with $t=|T/T_c-1|$. This result shows the strong infrared
sensitivity of the two-point correlator of the field $\chi$ at the
onset of chiral symmetry restoration. The detailed behavior of the
screening mass of the Polyakov loop near the phase transition
depends on the resummation procedure used to deal with the
infrared divergences.

The large $N$ framework motivated resummation \cite{Mocsy:2003un}
leads to:
\begin{eqnarray}
\Delta m^2_\chi(T)&=& -
\frac{2g_1^2(1+N_\pi)}{8\pi\,m_{\sigma}+(1+N_\pi)3\lambda }
 \, , \quad T>T_{\rm{c}\sigma} \\
\Delta m^2_\chi(T)&=& - \frac{2g_1^2}{8\pi\,M_{\sigma}+ 3\lambda}
 \, , \qquad \qquad \,\,\,\, T<T_{\rm{c}\sigma}\, .
\end{eqnarray}
This provides a qualitative improvement, since one expects that
the mass of the non-order parameter field remains finite at the
phase transition. From the above equations one finds that the
screening mass of the Polyakov loop is continuous and finite at
$T_{\rm{c}\sigma}$, and $\Delta
m_\chi^2(T_{\rm{c}\sigma})=-2g_1^2/(3\lambda)$, independent of
$N_\pi$, the number of pions. Even if the mass is not critical,
some associated quantities do display critical behavior. We define
the slope parameters for the singlet field as
\begin{eqnarray}
    {\cal D_\chi}^\pm &\equiv & \lim_{T\rightarrow T_{\rm{c}\sigma}^\pm}
    \frac{1}{\Delta m_\chi^2(T_{\rm{c}\sigma})}
\frac{d\, \Delta m_\chi^2(T)}{dT} \, . \label{slopes1}
\end{eqnarray}
These have the critical behavior ${\cal{D_\chi}}^{\pm}\sim
t^{{\nu}/{2}-1}$. However, as shown in \cite{Mocsy:2003un}
different critical exponents might emerge when one departs from
the large $N$ limit.

This analysis is not restricted to the chiral/deconfining phase
transition. The entanglement between the order parameter (the
chiral condensate) and the non-order parameter field (the Polyakov
loop) is universal.

%%%%%%%%%%%%%%%%%%%%%%%%%%%%%%%%
\section{Adjoint Representation}
\label{adjoint}

As a second application, consider two color QCD with two massless
Dirac quark flavors in the adjoint representation. Here the global
symmetry is $SU(2N_f)$ which breaks via a bilinear quark
condensate to $O(2N_f)$. The number of Goldstone bosons is
$2N_f^2+N_f-1$. We take $N_f=2$. There are two exact order
parameter fields: the chiral $\sigma$ field and the Polyakov loop
$\chi~$. Since the relevant interaction term $g_1\chi\sigma^2$ is
now forbidden, one might expect no efficient information transfer
between them. This naive statement is partially supported by
lattice data \cite{Karsch:1998qj}. While respecting general
expectations the following analysis suggests the presence of a new
and more elaborated structure which lattice data can clarify in
the near future.

The chiral part of the potential is given by (\ref{chiralpot})
with $2\,M=\sigma + i\,2\sqrt{2}\pi^a\,X^a$, $a=1,\dots,9$ and
$X^a\in {\cal A}(SU(4))-{\cal A}(O(4))$. $X^a$ are the generators
provided explicitly in equation (A.3) and (A.5) of
\cite{Appelquist:1999dq}. While the chiral part of the potential
takes the same form as for the fundamental representation there
are differences when expressing the potential in terms of the
component fields. These do not affect the following analysis. The
$Z_2$ symmetric potential for the Polyakov loop is
\begin{eqnarray}
V_\chi[\chi]=\frac{m_{0\chi}^2}{2}\chi^2+\frac{g_4}{4}\chi^4 \, ,
\end{eqnarray}
and the only interaction term allowed by symmetries is
\begin{equation}
V_{\rm{int}}[\chi,\sigma,\pi]=g_2\chi^2\,{\rm Tr }\left[M^{\dagger}M\right]
 =g_2\chi^2(\sigma^2+\pi^a\pi^a) \, .
\end{equation}
The effective Lagrangian has no knowledge of which transition, the
chiral or confinement, happens first. Although lattice data
already provides such information we find it instructive to
analyze separately all the possibilities.
\begin{figure}[t]
 \includegraphics[width=8.7 truecm, clip=true]{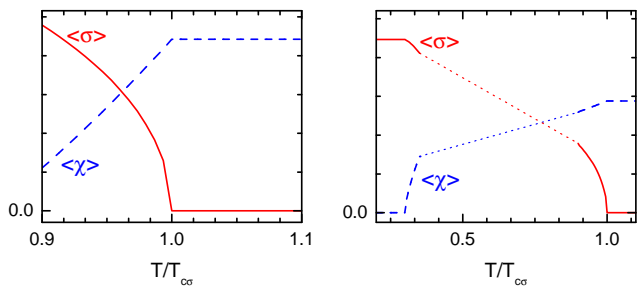}
\vspace{-1cm} \caption{Left panel: Behavior of the expectation
values of the Polyakov loop and chiral condensate close to the
chiral phase transition as a function of the temperature, with
quarks in the fundamental representation. Right panel: Same as in
left panel, for quarks in the adjoint representation and
$T_{\rm{c}\chi}\ll T_{\rm{c}\sigma}$ (see discussion in the
text).} \label{Figura1} \vspace{-.5cm}
\end{figure}

When chiral symmetry is restored before deconfinement
$T_{\rm{c}\sigma}\ll T_{\rm{c}\chi}~$ we consider three regimes:
For $T<T_{\rm{c}\sigma}$ the $Z_2$ symmetry is intact, while the
chiral symmetry is broken. Here $\langle\sigma\rangle
^2=-m^2/\lambda$. For $T>T_{\rm{c}\chi}$ the $Z_2$ is broken,
$\langle\chi\rangle^2=-m_{0\chi}^2/g_4$ and chiral symmetry is
restored. In both cases the coefficient of the relevant quadratic
term yielding condensation is not influenced by the expectation
values of the other field since the latter vanishes. In the
intermediate regime between the two critical temperatures both
symmetries are unbroken and
$\langle\sigma\rangle=\langle\chi\rangle=0$. In this intermediate
regime no trilinear interaction term between the fields is
induced. For $T<T_{\rm{c}\sigma}$ the interaction
$\langle\sigma\rangle\sigma\chi^2$, and for $T>T_{\rm{c}\chi}$ a
term  $\langle\chi\rangle\chi\sigma^2$ in the Lagrangian exists.
These interactions are innocuous for two reasons: i) They vanish
close to their respective phase transition, and ii) They cannot
induce any infrared divergent loops \cite{Mocsy:2003tr}. Thus for
$T_{\rm{c}\sigma}\ll T_{\rm{c}\chi}$ the two transitions are fully
separated, and neither of the two fields feels, even weakly, the
transition of the other.

The situation drastically changes when $T_{{\rm{c}}\chi}\ll
T_{{\rm{c}}\sigma}$. For $ T_{{\rm{c}}\chi}<T<T_{{\rm{c}}\sigma}$
both symmetries are broken, and the expectation values of the two
order parameter fields are linked to each other:
\begin{eqnarray}
\langle\sigma\rangle^2&=&-\frac{1}{\lambda}\left(m^2+
2g_2\langle\chi\rangle^2\right)\equiv
-\frac{m_\sigma^2}{\lambda}\, ,\nonumber
\\
\langle\chi\rangle^2&=&-\frac{1}{g_4}\left(m_{0\chi}^2+
2g_2\langle\sigma\rangle^2\right)\equiv -\frac{m_\chi^2}{g_4}\, .
\label{vevad}
\end{eqnarray}
The coupling $g_2$ is taken to be positive. One can show that
positivity of the square of the expectation values implies
$\lambda g_4-4g_2^2>0$. The latter is sufficient to make the
extremum of the potential a minimum. The expected behavior of
$m_\chi^2\sim (T-T_{\rm{c}\chi})^{\nu_{\chi}}$ and $m_\sigma^2\sim
(T-T_{\rm{c}\sigma})^{\nu_{\sigma}}$ near $T_{\rm{c}\chi}$ and
$T_{\rm{c}\sigma}$, respectively, combined with the result of eq.
(\ref{vevad}), yields in the neighborhood of these two transitions
the qualitative situation, illustrated in the right panel of
figure \ref{Figura1}. On both sides of $T_{\rm{c}\chi}$ the
relevant interaction term $g_2\langle \sigma\rangle\sigma\chi^2$
emerges, leading to a one-loop contribution to the static
two-point function of the $\sigma$ field $\propto \langle \sigma
\rangle^2 /m_\chi~$. Near the deconfinement transition
$m_\chi\rightarrow 0$ yielding an infrared sensitive screening
mass for $\sigma$. Similarly, on both sides of $T_{\rm{c}\sigma}$
the interaction term $\langle \chi\rangle\chi\sigma^2$ is
generated, leading to the infrared sensitive contribution $\propto
\langle \chi\rangle^2/m_\sigma$ to the $\chi$ two-point function.
We conclude, that when $T_{\rm{c}\chi}\ll T_{\rm{c}\sigma}$, the
two order parameter fields, a priori unrelated, do feel each other
near the respective phase transitions. It is important to
emphasize that the effective theory works only in the vicinity of
the two phase transitions. Interpolation through the intermediate
temperature range is shown by dotted lines in the right panel of
figure \ref{Figura1}. Possible structures here must be determined
via first principle lattice calculations.

The infrared sensitivity leads to a drop in the screening masses
of each field in the neighborhood of the transition of the other,
which becomes critical, namely of the $\sigma$ field close to
$T_{\rm{c}\chi}$, and of the $\chi$ field close to
$T_{\rm{c}\sigma}~$. These drops at the transition points are
expected, at the one-loop level, to behave as:
\begin{eqnarray}
\Delta m^2_\chi(T)&\sim&
-\frac{(g_2\langle\chi\rangle)^2}{|m_{\sigma}|}
 \sim t^{-\frac{\nu_{\sigma}}{2}} ,
\end{eqnarray}
and similarly, we have $ \Delta m^2_{\sigma}(T) \sim
t^{-{\nu_{\chi}}/{2}}$ near the $Z_2$ phase transition. In the
derivation of the above results we considered the expectation
values of the fields in the broken phases to be close to their
asymptotic values. The resummation procedure outlined in the
previous section predicts again a finite drop:
\begin{eqnarray}
\Delta m^2_\chi(T_{\rm{c}\sigma})=-\frac{8g_2^2\langle\chi\rangle
^2}{3\lambda}, \quad \Delta
m^2_{\sigma}(T_{\rm{c}\chi})=-\frac{8g_2^2\langle\sigma\rangle
^2}{3g_4}\, .
\end{eqnarray}
We thus predict the existence of substructures near these
transitions, when considering fermions in the adjoint
representation. Searching for such hidden behaviors in lattice
simulations would help to further understand the nature of phase
transitions in QCD.

%%%%%%%%%%%%%%%%%%%%
\section{Discussion}
\label{summary}

Via an effective Lagrangian approach we have seen how
deconfinement (i.e. a rise in the Polyakov loop) is a consequence
of chiral symmetry restoration in the presence of fermions in the
fundamental presentation. In nature quarks have small, but nonzero
masses, which makes chiral symmetry only approximate.
Nevertheless, the picture presented in this Letter still holds:
confinement is driven by the dynamics of the chiral transition.
The argument can be extended even further: If quark masses were
very large then chiral symmetry would be badly broken, and could
not be used to characterize the phase transition. But in such a
case the $Z_2$ symmetry becomes more exact, and by reversing the
roles of the protagonists in the previous discussion, we would
find that the $Z_2$ breaking drives the (approximate) restoration
of chiral symmetry. Which of the underlying symmetries demands and
which amends can be determined directly from the critical behavior
of the spatial correlators of hadrons or of the Polyakov loop
\cite{Mocsy:2003tr,{Mocsy:2003un}}.

With quarks in the adjoint representation we investigated two
scenarios. In a world in which chiral symmetry is restored first,
and then at some higher temperature deconfinement sets in,
$T_{\rm{c}\sigma}\ll T_{\rm{c}\chi}$, the two phase transitions
happen completely independent of each other. We know from
\cite{Karsch:1998qj} however, that $T_{\rm{c}\chi}\ll
T_{\rm{c}\sigma}$. In this case we have pointed to the existence
of an interesting structure, which was hidden until now: There are
still two distinct phase transitions, but since the fields are now
entangled, the transitions are not independent. This entanglement
is shown at the level of expectation values and spatial
correlators of the fields. More specifically, the spatial
correlator of the field which is not at its critical temperature
will in any case feel the phase transition measured by the other
field. Lattice simulations will play an important role in checking
these predictions.

The analysis can be extended for phase transitions driven by a
chemical potential. In fact, for two color QCD this is
straightforward to show. When considering fermions in the
pseudoreal representation there is a phase transition from a
quark-antiquark condensate to a diquark condensate
\cite{Hands:2001jn}. We hence predict, in two color QCD, that when
diquarks form for $\mu=m_{\pi}$, the Polyakov loop also feels the
presence of the phase transition exactly in the same manner as it
feels when considering the temperature driven phase transition.
Such a situation is supported by recent lattice simulations
\cite{Alles:2002st}. The results presented here are not limited to
describing the chiral/deconfining phase transition and can readily
be used to understand phase transitions sharing similar features.

Even if the effective Lagrangian approach a la Ginzburg--Landau is
an oversimplification, it allows on one hand to illuminate the
relevant physics involved, and on the other hand permits a
systematic study of different effects, such as a non-zero chemical
potential, quark masses, quark flavors and axial anomaly.

\acknowledgments We thank P.H. Damgaard, A.D. Jackson and R.
Pisarski for discussions. The work of F.S. is supported by the
Marie--Curie fellowship under contract MCFI-2001-00181.

%%%%%%%%%%%%%%%%%%

\end{document}